\documentclass[a4paper,prl,reprint,twocolumn,superscriptaddress]{revtex4-2}

\usepackage{ulem}
\usepackage{amssymb}
\usepackage{amsmath}
\usepackage{epsfig}
\usepackage{color}
\usepackage{graphics, graphicx}
\usepackage{bbold}
\usepackage{psfrag}
\usepackage{mathcomp}
\usepackage{amsmath}
\usepackage{amssymb}
\usepackage{mathrsfs}
\usepackage{subfigure}
\usepackage{verbatim}
\usepackage{xcolor,bm}
\usepackage[colorlinks,citecolor=blue,urlcolor=blue]{hyperref}

\def\de{\mathrm{d}}

\usepackage{mathrsfs}

\begin{document}
\title{Optical pumping through the Liouvillian skin effect}
\author{De-Huan Cai}
\affiliation{Hefei National Laboratory, University of Science and Technology of China, Hefei 230088,
China}
\author{Wei Yi}
\email{wyiz@ustc.edu.cn}
\affiliation{CAS Key Laboratory of Quantum Information, University of Science and Technology of China, Hefei 230026, China}
\affiliation{Anhui Province Key Laboratory of Quantum Network, University of Science and Technology of China, Hefei, 230026, China}
\affiliation{CAS Center For Excellence in Quantum Information and Quantum Physics, Hefei 230026, China}
\affiliation{Hefei National Laboratory, University of Science and Technology of China, Hefei 230088,
China}
\author{Chen-Xiao Dong}
\email{cxdong@hfnl.cn}
\affiliation{Hefei National Laboratory, University of Science and Technology of China, Hefei 230088,
China}

\begin{abstract}
The Liouvillian skin effect describes the boundary affinity of Liouvillian eignemodes that originates from the intrinsic non-Hermiticity of the Liouvillian superoperators. Dynamically, it manifests as directional flow in the transient dynamics, and the accumulation of population near open boundaries at long times.
Intriguingly, similar dynamic phenomena exist in the well-known process of optical pumping, where the system is driven into a desired state (or a dark-state subspace) through the interplay of dissipation and optical drive.
In this work, we show that typical optical pumping processes can indeed be understood in terms of the Liouvillian skin effect.
By studying the Liouvillian spectra under different boundary conditions, we reveal that the Liouvillian spectra of the driven-dissipative pumping process sensitively depend on the boundary conditions in the state space, a signature that lies at the origin of the Liouvillian skin effect.
Such a connection provides insights and practical means for designing efficient optical-pumping schemes through engineering Liouvillian gaps under the open-boundary condition.
Based on these understandings, we show that the efficiency of a typical side-band cooling scheme for trapped ions can be dramatically enhanced
by introducing counterintuitive dissipative channels.
Our results provide a useful perspective for optical pumping, with interesting implications for state preparation and cooling.
\end{abstract}

\maketitle

{\it Introduction.---}
Optical pumping is a fundamentally important technique in the study of atomic, molecular, and optical physics~\cite{Franzen,Kastler,Happer1,Happer2,Happer3,Walker,Appelt,Han,Zare,Broyer,Viteau,Balling,Olsen,Pitz}.
Originally developed to achieve population inversion necessary for lasing~\cite{Kastler,Happer1}, it has become the standard practice to cyclically pump atoms to a given quantum state~\cite{Franzen,Kastler,Happer1,Jau,Weber}, often with a well-defined magnetic quantum number. More generally, through the ingenious design of optical drive and dissipation,
a quantum open system can be driven into a desired steady state (or a desired dark-state subspace) at long times~\cite{Zoller1,Zoller2,Zoller3,Zoller4,Zhang,Lin,Wang}.
Such general optical pumping processes are widely used for state preparation and cooling~\cite{Zhang,Wineland1,Wineland2,Wineland3,Wineland4,Wineland}, and offer promising paradigms for quantum simulation with atoms~\cite{Zoller1,Zoller2,Zoller3,Zoller4}.
Phenomenologically, a typical optical pumping process manifests two salient features, the directional flow of population in the state space, and the long-time population accumulation in the final steady state, which, given its dark-state nature, can be considered as a boundary in the state space.
Intriguingly, these features also manifest in systems with the non-Hermitian skin effect, a phenomenon that has attracted extensive interest in recent years~\cite{nhsetheory1,nhsetheory4,nhsetheory5,nhsetheory7,nhsetheory8,nhsetheory9,nhsetheory10,nhsetheory2,nhsetheory3,
nhsetheory6,nhsetheory11,nhsetheory12,nhsetheory13,nhsetheory14,NHSEdynamic1,NHSEdynamic2,NHSEdynamic3,NHSEdynamic4,NHSEdynamic5,
nhseexperiment2,nhseexperiment4,nhseexperiment7,nhseexperiment8,nhseexperiment9,nhseexperiment1,nhseexperiment5,nhseexperiment12}.

The non-Hermitian skin effect describes the accumulation of eigenstates near the boundaries of certain non-Hermitian systems~\cite{nhsetheory1,nhsetheory2,nhsetheory3,nhsetheory4,nhsetheory5,nhsetheory6,nhsetheory7,nhsetheory8,nhsetheory9,
nhsetheory10,nhsetheory11,nhsetheory12,nhsetheory13,nhsetheory14}.
It derives from the instability of the eigenvalue problems of non-Hermitian matrices to boundary perturbations, and has profound impact on the band and spectral topologies~\cite{nhsetheory11,nhsetheory12,nhsetheory14}, as well as the bulk dynamics~\cite{NHSEdynamic1,NHSEdynamic2,NHSEdynamic3,NHSEdynamic4,NHSEdynamic5,nhsetheory8,nhsetheory9}.
Experimentally, the non-Hermitian skin effect and its various manifestations have been observed in classical systems with gain and/or loss~\cite{nhseexperiment2,nhseexperiment4,nhseexperiment7,nhseexperiment8,nhseexperiment9}, and in the conditional dynamics of quantum open systems subject to post selection~\cite{nhseexperiment1,nhseexperiment5,nhseexperiment12}.
But the non-Hermitian skin effect also arises in the full-fledged quantum dynamics governed by the Lindblad master equation, wherein the Liouvillian superoperator can be mapped to a non-Hermitian matrix in an enlarged Hilbert space.
Alternatively, under the master equation, the single-particle correlation evolves according to a non-Hermitian damping matrix~\cite{NHSEdynamic4,Li}.
The corresponding non-Hermitian skin effect in quantum open systems, dubbed the Liouvillian skin effect~\cite{lse1}, hosts chiral damping and directional bulk flow in the transient dynamics, as well as various boundary-sensitive long-time behaviors, such as the time scale at which the steady state is approached, and the boundary affinity of steady-state population~\cite{Wang,lse1,lse2,lse4,lse5,lse7}.
While the Liouvillian skin effect has yet to be explicitly demonstrated in experiments, the resemblance of its dynamic consequences with those of optical pumping strongly suggests an intimate, if not direct, connection between them.

\begin{figure}[tbp]
  \centering
  \includegraphics[width=8.5cm]{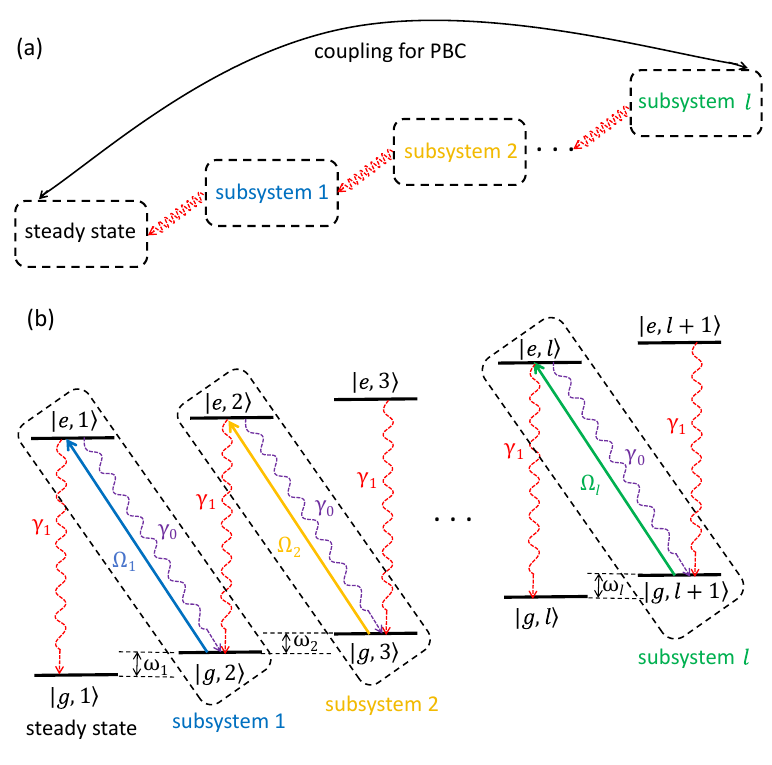}
  \caption{(a) Schematic illustration of a general optical pumping process, where a series of subsystems are connected by directional dissipation ending in a steady state.
  Both coherent and incoherent couplings exist between quantum states within each subsystem.
  (b) A typical example of the general scheme in (a), where $|g\rangle$ and $|e\rangle$ are the electronic ground and excited states, and $|l\rangle$ are the states in the ground- and excited-state manifolds, with energy offsets $\omega_l$ between adjacent states in the same manifold.
  The purple and red dashed arrows indicate the spontaneous decay from the excited state $|e,l\rangle$ to the ground states $|g,l+1\rangle$ and $|g,l\rangle$, respectively.
  The solid arrows with different colors indicate the resonant optical couplings between states in the ground-state manifold and the excited states. The effective Rabi coupling rates $\Omega_l$ and the decay rates $\gamma_{0,1}$ are also illustrated. }
  \label{fig:fig1}
\end{figure}

In this work, we show that typical optical pumping processes can indeed be understood in terms of the Liouvillian skin effect of the underlying quantum master equation.
As illustrated in Fig.~\ref{fig:fig1}(a), we focus on a generic optical pumping setup, where a series of otherwise independent quantum-state sectors (labeled by $l$) are connected by directional dissipation.
The quantum states within each sector are coupled by coherent optical fields, and may subject to additional incoherent dissipative processes in between.
A discrete translational symmetry in $l$ is possible, but not necessary.
Typical examples of such a general setup include the simplest optical pumping process in a three-level system, and the side-band cooling in trapped ions.
In these examples, an open boundary condition (OBC) is naturally present, with the final state of the pumping process forming an open boundary.
However, for the sake of discussion, a formal periodic boundary condition (PBC) can also be enforced by connecting the left-most and right-most sectors [as illustrated in Fig.~\ref{fig:fig1}(a)].
Note that the formal PBC here does not imply lattice translational symmetry, which is itself not necessary for the onset of the non-Hermitian skin effect ~\cite{hwLi,NHSEdynamic3}.
We take a typical side-band cooling configuration as an example, and study the Liouvillian spectra of the system.
We find that the eigenspectra sensitively depend on the boundary conditions, a signature that lies at the origin of the Liouvillian (or non-Hermitian) skin effect.
The existence of the Liouvillian skin effect is further confirmed by the directional bulk flow and the accumulation of the steady-state population at the open boundary, both of which are also natural consequences of the side-band cooling (or optical pumping) setup.

Such a connection provides insights on the further design of efficient optical pumping schemes.
Specifically, since the time for the system to reach the steady state is determined by the Liouvillian gap, the efficiency of the optical pumping process can be enhanced by engineering larger Liouvillian gaps.
Through analytic and numerical analyses, we identify the condition to maximize the Liouvillian gap of our system, which is surprisingly achieved by introducing dissipative processes that are opposite in direction to the bulk flow.

\begin{figure*}[tbp]
  \centering
  \includegraphics[width=13.5cm]{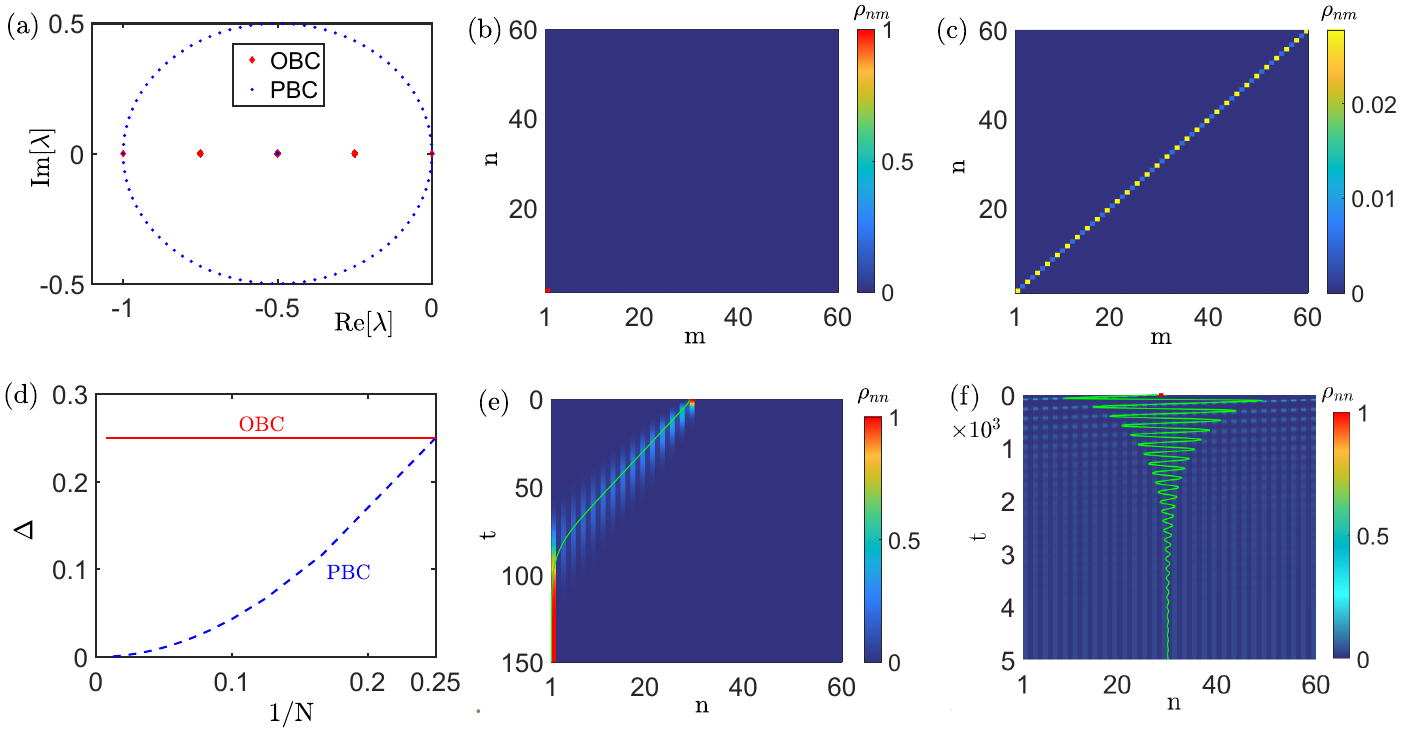}
  \caption{Liouvillian spectrum and density-matrix dynamics. (a) The red square dots and blue dots denote the eigenvalues of Liouvillian superoperator $\mathcal{L}$ in OBC and PBC, respectively. (b) and (c) are the eigenmodes of zero eigenvalue for Liouvillian superoperator $\mathcal{L}$ under OBC and PBC, respectively. (d) The Liouvillian gap as a function of the size system, for different boundary conditions. (e) and (f) are the time evolution of distribution for eigenmodes under OBC and PBC respectively, with the initial state $|g,15\rangle=1$. The green solid line indicates the time evolution of the average
value for the energy state's index, i.e., $\langle n\rangle = \sum_{n}n\rho_{nn}$. The dimension of the Hilbert space of the system is $N=60$ except for (d). Other parameters: $\Omega=0.25$,$\gamma_1$=1.}
  \label{fig:fig2}
\end{figure*}

{\it Liouvillian skin effect in optical pumping.---}
As illustrated in Fig.~\ref{fig:fig1}(b), we consider a concrete example of the general optical pumping process, where external light fields couple transitions from the ground to the excited states, and, aided by dissipative processes, eventually pump the system to a given steady state.
Specifically, a set of ground states with energy intervals $\{\omega_l\}$($l=1,2...$) are labeled as $\{|g,l\rangle=|n=2l-1\rangle\}$, and the corresponding excited states are labeled as $\{|e,l\rangle=|n=2l\rangle\}$.
The Rabi frequencies of the coherent optical couplings are $\{\Omega_l\}$, and $\gamma_0$ and $\gamma_1$ are the decay rates from an excited state to different states in the ground-state manifold.
Physically, $l$ can label magnetic quantum numbers in the ground- and excited-state hyperfine manifolds~\cite{Jau}, in which case the scheme in Fig.~\ref{fig:fig1}(b) corresponds to a typical optical pumping for state preparation. Alternatively, $l$ can label phonon side bands in trapped ions, in which case Fig.~\ref{fig:fig1}(b) depicts side-band cooling~\cite{Zhang,Wineland1,Wineland2,Wineland3,Wineland4,Wineland}.
Regardless of the physical correspondence, the time evolution of the density matrix
under the couplings of Fig.~\ref{fig:fig1}(b) is determined by the Lindblad master equation (we take $\hbar=1$)~\cite{section1,section2}
\begin{align}
\frac{\de\rho}{\de t} = -i[H,\rho] + \sum_{l,p}(2L_{l,p}\rho L^{\dag}_{l,p}- \{L^{\dag}_{l,p},L_{l,p}\}\rho) \equiv \mathcal{L}(\rho). \label{eq:Lindblad}
\end{align}
Here the coherent Hamiltonian reads
\begin{align}
 H = & \sum_{l}\Big[(\sum_{j=1}^{l}\omega_j)(|g,l+1\rangle\langle g,l+1| + |e,l+1\rangle\langle e,l+1|)\Big]\nonumber\\
 &+ \sum_{l}\Omega_{l}(|e,l\rangle\langle g,l+1| + \mathrm{H.c.}), \label{eq:Hamiltonian}
\end{align}
and the quantum jump operators are
\begin{align}
L_{l,p=0}=\sqrt{\gamma_0}|g,l+1\rangle\langle e,l|,\quad L_{l,p=1}=\sqrt{\gamma_1}|g,l\rangle\langle e,l|. \label{eq:Lindbladoper}
\end{align}

We denote the Hilbert-space dimension of the system as $N$, with $n_{max}=2l_{max}=N$. Then the right and left eigenmodes of the Liouvillian superoperator $\mathcal{L}$, defined in an $N^2$-dimensional extended Hilbert space, are given by
\begin{align}
\mathcal{L}(\rho_{\mu}^{R}) = \lambda_{\mu}\rho_{\mu}^{R}, \quad \quad \mathcal{L}^{\dag}(\rho_{\mu}^{L}) = \lambda_{\mu}^{\ast}\rho_{\mu}^{L}, \label{eq:Liouvillian}
\end{align}
with $\mu=1,2,3,...,N^2$. The right and left eigenmodes are normalized as $\sqrt{\langle\rho^{R}_{\mu}|\rho^{R}_{\mu}\rangle}=\sqrt{\langle\rho^{L}_{\mu}|\rho^{L}_{\mu}\rangle}=1$, and are orthogonal to each other ($\sqrt{\langle\rho^{L}_{\mu}|\rho^{R}_{\nu}\rangle}=0$) when their eigenvalues are different ($\lambda_{\mu}\neq\lambda_{\nu}$).
In particular, the eigenmodes of $\mathcal{L}$ with vanishing eigenvalues are the steady states of the system, with $\mathcal{L}(\rho_{ss}) =0$.

It follows that the density matrix of the initial state can be expanded as
\begin{align}
\rho_{\mathrm{ini}}=\sum_{\mu=1}^{N^2}c_{\mu}\rho_{\mu}^{R}, \label{eq:initial}
\end{align}
where $c_{\mu}=\langle\rho^{L}_{\mu}|\rho_{\mathrm{ini}}\rangle/\langle\rho^{L}_{\mu}|\rho^{R}_{\mu}\rangle$ according to the completeness condition $\sum_{\mu}|\rho^{R}_{\mu}\rangle\langle\rho^{L}_{\mu}|/\langle\rho^{L}_{\mu}|\rho^{R}_{\mu}\rangle=1$.
Thus, the time evolution of the density matrix can be written as
\begin{align}
\rho(t)=\sum_{\mu=1}^{N^2}c_{\mu}e^{\lambda_{\mu}t}\rho_{\mu}^{R}. \label{finalstate1}
\end{align}
Note that the real parts of the eigenvalues of the excited eigenmodes (those that are not steady states)
must be negative to ensure that their contributions in Eq.~\ref{finalstate1} would be exponentially small after a long enough time evolution, as the system approaches the steady states. Here we set $\rho_{ss}=\rho_{\mu=1}^{R}$, and assume that all eigenvalues are indexed in descending order according to their real parts: $0=\lambda_1>\mathrm{Re}[\lambda_2]\geq\mathrm{Re}[\lambda_3]...\geq\mathrm{Re}[\lambda_{N^2}]$. Equation \ref{finalstate1} can then be rewritten as
\begin{align}
\rho(t)=\rho_{ss} + \sum_{\mu=2}^{N^2}c_{\mu}e^{\lambda_{\mu}t}\rho_{\mu}^{R}. \label{eq:evolution}
\end{align}
Importantly, the Liouvillian gap is defined as $\Delta=|\mathrm{Re}[\lambda_2]|$, which describes the asymptotic decay rate of the system toward the steady states at long times~\cite{section3}.

We first consider the simple case with $\omega_l = 0$, $\Omega_l = \Omega$, and $\gamma_0 =0$.
It follows that Hamiltonian (\ref{eq:Hamiltonian}) is simplified to $H=\sum_{l}\Omega(|e,l\rangle\langle g,l+1| + \mathrm{H.c.})$, and only a single quantum jump process exists for each pair of ground and excited states, given by $L_{l,1}$. In Hamiltonian (\ref{eq:Hamiltonian}), states with the smallest and largest $n$ indices are not coupled. This corresponds to an OBC in the state space. By contrast, one may consider an artificial PBC, where all states are cyclically coupled. Such a PBC is achieved by adding the term $(\Omega|e,l_{\mathrm{max}}\rangle\langle g,1|  +\mathrm{H.c.})$ to Eq.~(\ref{eq:Hamiltonian}), where $l_{\mathrm{max}}$ is the maximum $l$. Although the PBC is unphysical, it offers insights to the setup as we detail below. Alternatively, one may consider the state label $n$ as lattice sites along a synthetic dimension. Different boundary conditions in the synthetic dimension then directly correspond to those in the state space. With these understandings, we now study the Liouvillian spectrum and dynamic evolution of the master equation.

As depicted in Fig.~\ref{fig:fig2}(a), the eigenvalues of the Liouvillian superoperator $\mathcal{L}$ under the PBC form a closed loop on the complex plane, enclosing those under the OBC.
This is reminiscent of the spectral topology of non-Hermitian Hamiltonians with the skin effect, and is an outstanding signature for the Liouvillian skin effect.
In either case, the drastic difference in the eigenspectrum under different boundary conditions originates from the instability of non-Hermitian matrices to boundary perturbations.
Fig. ~\ref{fig:fig2}(b) shows the density-matrix elements $\rho_{nm}$ of the steady state under the OBC.
Here the density-matrix element is defined as $\rho_{nm}=\langle n|\rho_{\mu=1}^{R}|m\rangle$.
The steady state is indeed localized in $|g,l=1\rangle$, corresponding to an open boundary.
The corresponding steady state under the PBC is shown in Fig.~\ref{fig:fig2}(c), where uniform distributions in $l$ are observed for both the ground and excited states.
A closer look reveals that, in the steady state under the PBC, the majority of the population is in the ground state.

Another drastic distinction between the Liouvillian spectrum under OBC and PBC is the Liouvillian gap.
As shown in Fig.~\ref{fig:fig2}(d), the Liouvillian gap $\Delta$ tends to zero as the size of the system increases under the PBC.
By contrast, the gap is independent of the system size under the OBC.
A finite Liouvillian gap implies that the density matrix in Eq.~(\ref{eq:evolution}) converges exponentially fast to the steady state at long times. Whereas a vanishing Liouvillian gap implies an algebraic convergence, such that the relaxation time diverges for $\Delta\rightarrow 0$~\cite{section4}.

Taking the size of the system as $N=60$ in Fig.~\ref{fig:fig2}(e) and (f), we evolve the system according to Eq.~(\ref{eq:Lindblad}), while setting the initial state to $|g,15\rangle=|n=29\rangle$. Under the OBC, the occupation rapidly flows toward the boundary and eventually evolves to the steady state as shown in Fig.~\ref{fig:fig2}(b). This is the dynamic signature of the Liouvillian skin effect.
In the context of optical pumping, such a directional flow is the underlying mechanism for state preparation and cooling.
For instance, in trapped ions, the index $l$ corresponds to the phonon modes. The coherent optical drives are implemented by side-band couplings, and the directional flow toward $l=0$ corresponds to cooling of the external ion motion. The timescale or efficiency of the cooling process is then determined by the Liouvillian gap under the OBC.
Under the PBC, since the Liouvillian gap is much smaller, the time it takes to relax to the steady state is much longer, and diverges in the thermodynamic limit.
We note that the conclusions above also hold for the more general cases of $\omega_l \neq 0$ and state-dependent $\Omega_l$~\cite{supple}.



Furthermore, one can show that the general optical pumping process illustrated in Fig.~\ref{fig:fig1} leads to a block-diagonal structure of the Liouvillian superoperator $\mathcal{L}$ (related to the divisibility of the setup into subsystems), so that we can analytically solve the Liouvillian spectrum and the Liouvillian gap~\cite{supple}.
We find that the Liouvillian spectra are always distinct under the PBC and OBC, with different asymptotic behaviors of the Liouvillian gap under different boundary conditions, as the system approaches the thermodynamic limit.
This phenomenon is referred to as the Liouvillian skin effect, which is analogous to the non-Hermitian skin effect observed in non-Hermitian lattice models.
Physically, the Liouvillian skin effect in our system arises from the divisibility mentioned above, and the non-reciprocal quantum jump operators~\cite{supple}.

\begin{figure}[tbp]
  \centering
  \includegraphics[width=8.5cm]{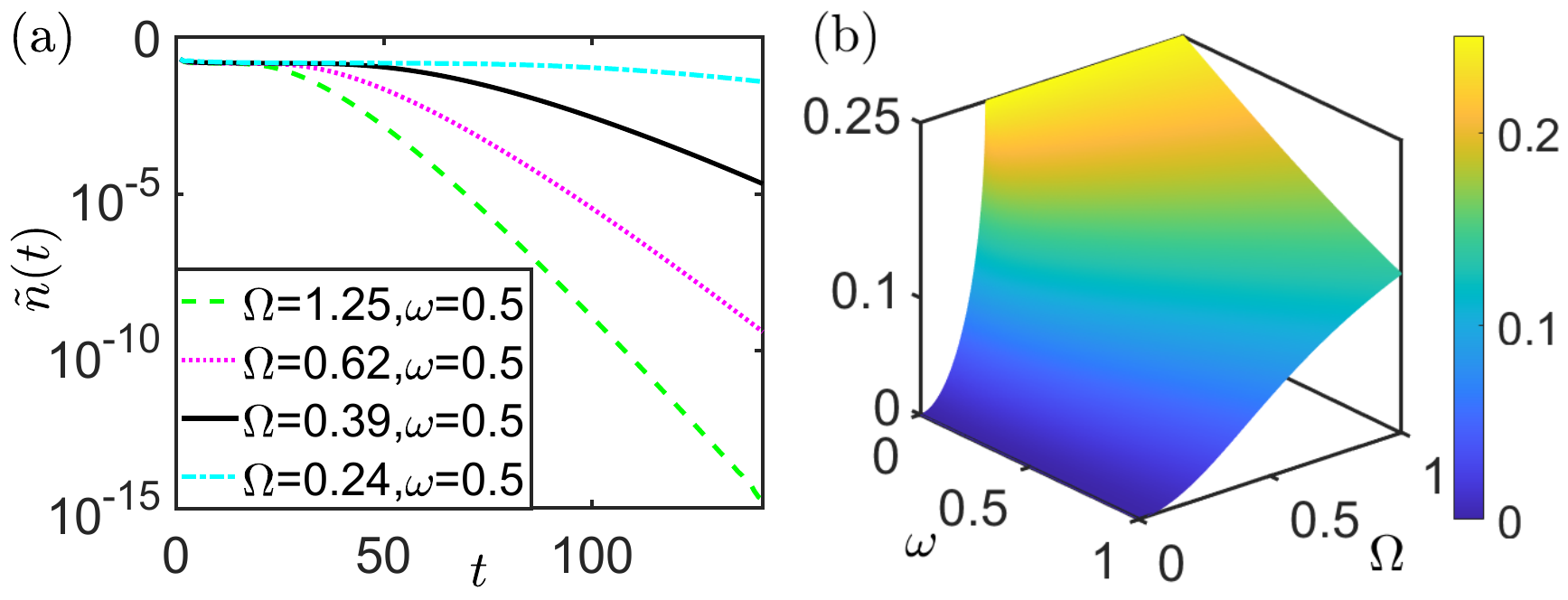}
  \caption{Liouvillian gap and the long-time damping dynamic.
  (a) The long-time damping follows an exponential law after an initial power law stage, where the system size is $N=50$.
  (b) The Liouvillian gap as a function of energy interval $\omega$ and Rabi frequency $\Omega$ with $N=50$.
  We perform numerical calculations with OBC, the decay rates are $\gamma_0=0$ and $\gamma_1=1$.}
  \label{fig:fig3}
\end{figure}

%

{\it Designing efficient pumping scheme.---}
Facilitated by the understandings above, we now show that the pumping scheme in Fig.~\ref{fig:fig1}(b) can be optimized.
Here we set $\omega_l = \omega$ and $\Omega_l = \Omega$ to simplify discussions, but our results qualitatively hold for schemes without the translational symmetry. The latter can be important for side-band cooling in trapped ions in the Lamb-Dicke regime, where the coupling strength between different side bands scale as $\sqrt{n}$~\cite{Lin}.

In our system, any initial state evolves towards a steady state. To quantify the damping dynamics, we calculate the particle-number deviation from that of the steady state, defined as $\tilde{n}(t)=\text{Tr}[\rho(t)-\rho(t\to \infty)]$. As shown in Fig.~\ref{fig:fig3}(a), the damping of $\tilde{n}(t)$ depends on the initial state and the Liouvillian gap. With the same initial states, the damping dynamic accelerates when the Liouvillian gap increases.

Generally, in the experiments, the energy offset $\omega$ is smaller than the Rabi frequency $\Omega$. Specifically,  when $\omega=0$, the Liouvillian gap increases with $\Omega/\gamma_1$ when $\Omega/\gamma_1<1/4$, reaching a maximum of $\gamma_1/4$ when $\Omega/\gamma_1>1/4$, as illustrated in Fig.~\ref{fig:fig3}(b). Subsequently, if $\omega\neq0$, the Liouvillian gap consistently decreases with increasing $\omega$.
As a result, the maximum possible Liouvillian gap is $\gamma_1/4$ when $\gamma_0=0$, which is consistent with previous studies~\cite{Zhang}.
In the following, we aim to further increase the Liouvillian gap by introducing new decay channels.

\begin{figure}[tbp]
  \centering
  \includegraphics[width=8.5cm]{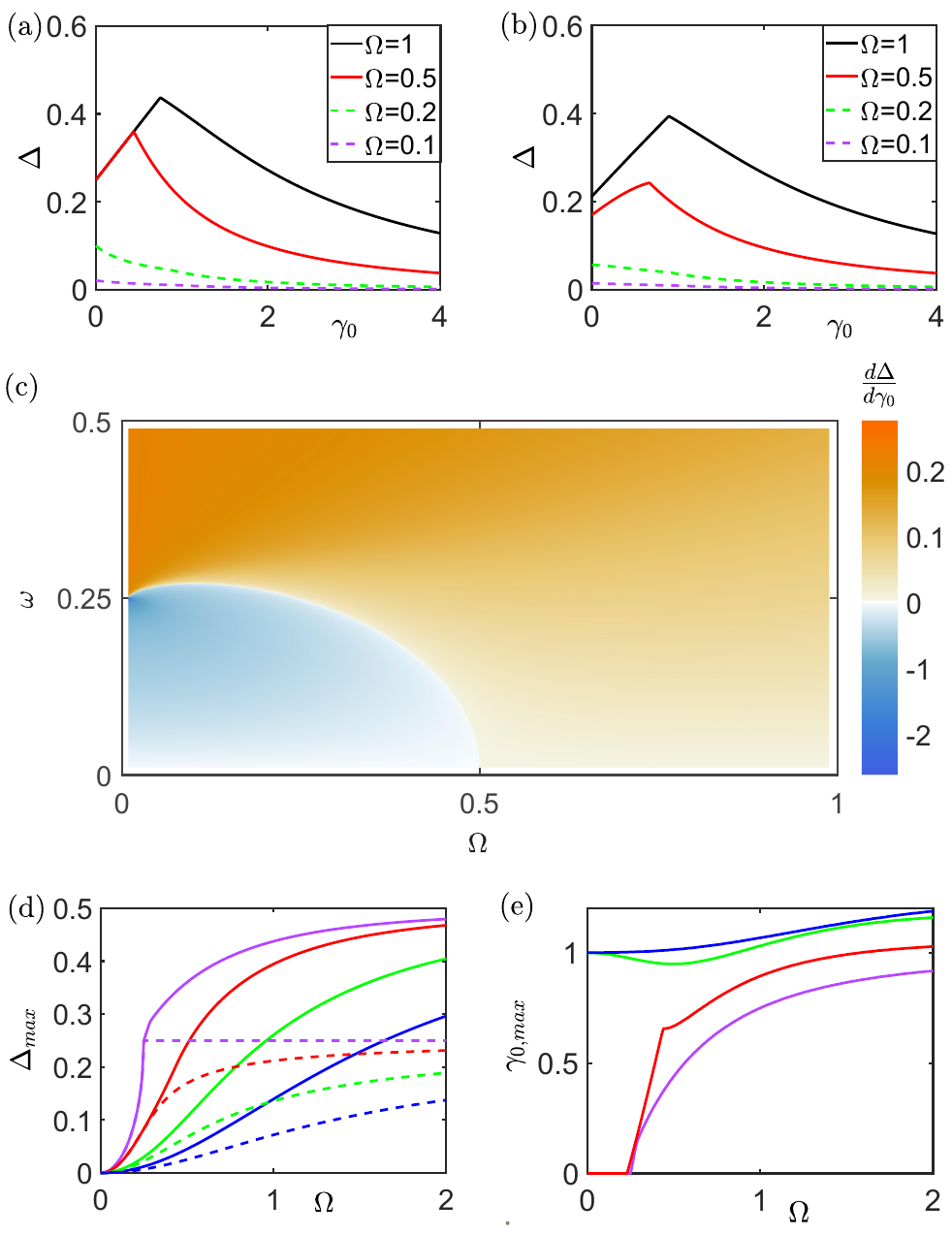}
  \caption{Liouvillian gap for $\gamma_0 \neq 0$.
(a) and (b) show the Liouvillian gap with increasing $\gamma_0$ for different $\Omega$, for $\omega =0$ and $0.3$, respectively. (c) The change rate of the Liouvillian gap at $\gamma_0$ ($\left.\de \Delta/\de \gamma_0\right|_{\gamma_0=0}$).
(d) and (e) show the maximum Liouvillian gap and the optimal decay rate $\gamma_{0,\text{max}}$ for the maximum Liouvillian gap, under different $\omega$ and $\Omega$. The magenta, red, green and blue lines (include solid and dashed lines) in (d) and (e) correspond to the parameters of $\omega=0$, $\omega=0.3$, $\omega=1$, and $\omega=2$, respectively. In (d), dashed lines represent the Liouvillian gap when $\gamma_0 = 0$,  solid lines represent the maximum Liouvillian gap for $\gamma_0\neq 0$. For all the plots, we set $\gamma_1=1$.
	}
  \label{fig:fig4}
\end{figure}


We first introduce an additional decay term given by the jump operator
\begin{equation}
L_{l,p=2}=\sqrt{\gamma_{2}}|g,l\rangle\langle e,l+1|,
\end{equation}
which enhances the dissipation in the direction of the steady state.
While such a term does not change the discussion on the Liouvillian superoperator under OBC, it contributes to an increased decay rate within the subsystem, effectively transforming $\gamma_1$ to $\gamma_1+\gamma_{2}$.
Consequently, the maximum Liouvillian gap becomes $\gamma_1/4+\gamma_{2}/4$, and the pumping efficiency is enhanced.
Likewise, we can introduce longer-distance decay terms to similar effects.

Alternatively, we consider the decay term $L_{l,0}$, leading to transitions within each subsystem. As illustrated in Fig.~\ref{fig:fig1}(b), the direction of the dissipation is opposite to that of the directional flow toward the steady state.
From numerical calculations, we identify two distinct behaviors of the Liouvillian gap when varying $\gamma_0$.
First, the Liouvillian gap monotonically decreases to 0 with increasing $\gamma_0$, shown as dashed lines in Fig.~\ref{fig:fig4}(a)(b). Second, the Liouvillian gap increases to a maximum value before decreasing to 0, shown as solid lines in
Fig.~\ref{fig:fig4}(a)(b).  Here we use $\left.\de \Delta/\de \gamma_0\right|_{\gamma_0=0}$ to differentiate the parameter regimes for these different behaviors, as shown in Fig.~\ref{fig:fig4}(c). When $ \left.\de \Delta/\de \gamma_0\right|_{\gamma_0=0}<0$, the Liouvillian gap monotonically decreases with increasing $\gamma_0$; otherwise, the Liouvillian gap increases to a maximum value before decreasing to 0, resulting in a larger Liouvillian gap for appropriate values of $\gamma_0$ compared to the case where $\gamma_0=0$.
We then numerically calculate the maximum Liouvillian gap for different $\omega$ and $\Omega$, as shown in Fig.~\ref{fig:fig4}(d)(e).
In general, the maximum Liouvillian gap increases with larger $\Omega$ and smaller $\omega$.
The optimal decay rate $\gamma_{0,\text{max}}$ for achieving the maximum Liouvillian gap shows intricate behavior in conjunction with other parameters. Introducing the decay term $L_{l,0}$ yields a potential maximum Liouvillian gap of $\gamma_1/2$, achievable under the parameters $\Omega\rightarrow \infty$, $\omega=0$, and $\gamma_1=\gamma_0$.

{\it Summary.---}
To summarize, we show that typical optical pumping processes can be understood from the perspective of the Liouvillian skin effect. We confirm this understanding through the Liouvillian eigenspectrum and open-system dynamics for a concrete optical pumping setup involving coherent optical drives and directional dissipation.
We further illustrate that such an understanding provides means to optimize the pumping efficiency.
Our results are helpful for state preparation and cooling in quantum simulation and computation where optical pumping is inevitable.
Finally, while our discussion focuses on typical instances of optical pumping and side-band cooling, our analysis can also be applied to similar processes of spontaneous spin polarization~\cite{Minganti}, and many-body quantum-state preparation through reservoir engineering~\cite{Diehl,Cho}, where the final steady state can be understood as a boundary in the synthetic state space.

\begin{acknowledgments}
This work is supported by the National Natural Science Foundation of China (Grant No. 12374479), and by the Innovation Program for Quantum Science and Technology (Grant No. 2021ZD0301205, 2021ZD0301904).
\end{acknowledgments}

\end{document}


\title{Supplemental Material for ``Optical pumping through the Liouvillian skin effect''}
\author{De-Huan Cai}

\author{Wei Yi}
\email{wyiz@ustc.edu.cn}

\author{Chen-Xiao Dong}
\email{cxdong@hfnl.cn}

\affiliation{Hefei National Laboratory, University of Science and Technology of China, Hefei 230088,
China}
\affiliation{CAS Key Laboratory of Quantum Information, University of Science and Technology of China, Hefei 230026, China}
\affiliation{Anhui Province Key Laboratory of Quantum Network, University of Science and Technology of China, Hefei, 230026, China}
\affiliation{CAS Center For Excellence in Quantum Information and Quantum Physics, Hefei 230026, China}
\affiliation{Hefei National Laboratory, University of Science and Technology of China, Hefei 230088,
China}

\affiliation{Hefei National Laboratory, University of Science and Technology of China, Hefei 230088,
China}

\date{\today}


\maketitle

\onecolumngrid

\section{SI.~The Liouvillian spectrum and density-matrix dynamics under the OBC and the PBC}
\begin{figure*}[hbp]
  \centering
  \includegraphics[width=9cm]{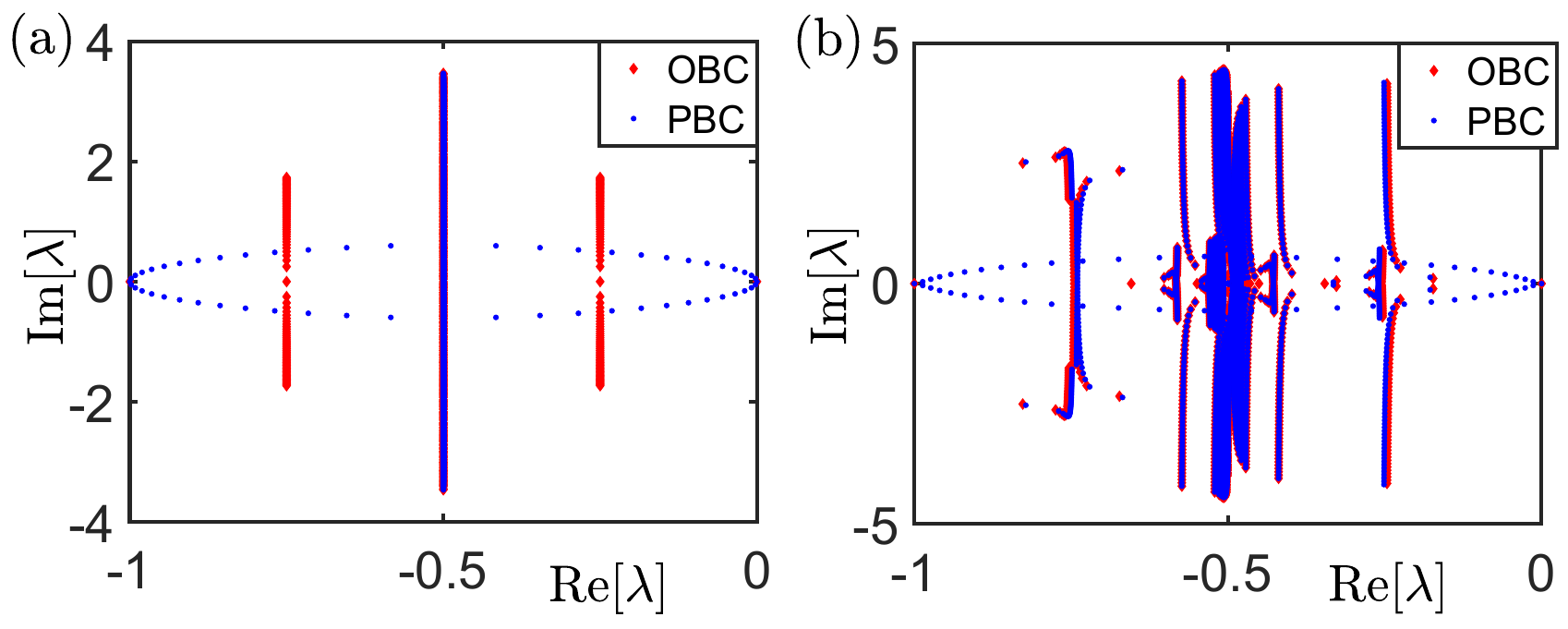}
  \caption{
 The Liouvillian spectrum under the OBC (red diamond) and PBC (blue dots), respectively.
(a) We take $\Omega_l=\sqrt{l}\Omega$, $\omega=0$. (b) We take $\Omega_l=\sqrt{l}\Omega$, $\omega=0.05$. Other parameters are $\Omega=0.25$, $\gamma_1=1$, $\gamma_0=0$, with the system size is $N=100$.}
  \label{fig:figS1}
\end{figure*}

In Fig.~2(a) of the main text, we obtain the Liouvillian spectrum under the PBC, which is a closed loop on the complex plane, enclosing the eigenspectrum under the OBC. For the calculation, we take the parameters: $\Omega_{l}=\Omega$, $\Omega=0.25$, and $\gamma_1=1$.
For the more general cases, we have $\omega_l \neq 0$, and state-dependent $\Omega_l$ (but still with $\gamma_0 =0$).
As shown in Fig.~\ref{fig:figS1}, now the Liouvillian spectrum under the PBC no longer encloses the one under the OBC.
While the complicated spectral features under the PBC in Fig.~\ref{fig:figS1}(b) are a direct result of finite $\omega_l$, a closed loop is still present for $\omega_l
\neq 0$.
In these cases, the Liouvillian skin effect persists, and the steady state under the OBC remains the same as that in Fig.~2(b) of the main text. The long-time evolution of the system generates a directional flow, similar to the results shown in Fig.~2(e) of the main text, and the relaxation time depends on the Liouvillian gap.

\begin{figure*}
  \centering
  \includegraphics[width=14cm]{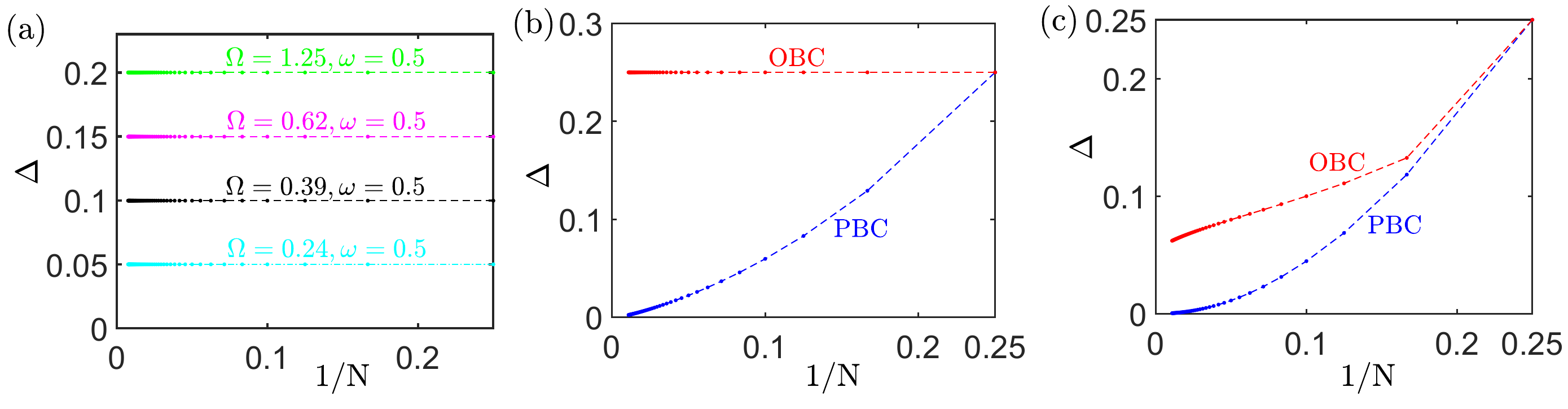}
  \caption{Liouvillian gap as a function of the system size.
(a) We take $\omega_{l}=\omega$ and $\Omega_l=\Omega$. The Liouvillian gaps are independent of the dimension of the system in OBC, where different Rabi frequencies and energy intervals result in different gaps; (b) We take $\Omega_l=\sqrt{l}\Omega$; (c) We take  $\Omega_{l}=\Omega_{0}+(\Omega-\Omega_{0})/\sqrt{l}$, $\Omega_{0}=0.15$. Other parameters: $\Omega=0.25,\gamma_1=1$.}
  \label{fig:figS2}
\end{figure*}

On the other hand, as shown in Fig.~2(d)  of the main text, the OBC gap is independent on the system size $N$ (Here, we take the parameters: $\Omega_{l}=\Omega$, $\Omega=0.25$, and $\gamma_1=1$). Similar behavior is observed in Fig.~\ref{fig:figS2}(a), where we keep the coupling pattern ($\Omega_{l}=\Omega$) unchanged, and vary $\Omega$ and $\omega$. 
We then modify the effective Rabi-coupling rates in Fig.~2(d) of the main text to $\Omega_{l}=\sqrt{l}\Omega$,
which is the typical coupling pattern for an ion-trap side-band cooling setup in the Lamb-Dicke limit. 
As shown in Fig.~\ref{fig:figS2}(b), the OBC gap also remains independent of the system size, while the PBC gap tends to zero for large $N$, similar to the case in Fig.~2(d) of the main text. 
Finally, we take $\Omega_{l}=\Omega_{0}+(\Omega-\Omega_{0})/\sqrt{l}$, which is not physical but produces a different scenario. As shown in Fig.~\ref{fig:figS2}(c), for $N\to \infty$, the PBC gap still vanishes, but the OBC gap approaches a finite value.

Therefore, we believe it is quite general that the asymptotic behaviors of the Liouvillian gap are different under distinct boundary conditions, as the system approaches the thermodynamic limit. We also emphasize that, while other coupling patterns may exist, the most significant cases are $\Omega_{l} = \Omega$ and $\Omega_{l} =\sqrt{l} \Omega$, as they are directly relevant to physical systems. In the next section, we will illustrate the structure of the Liouvillian spectrum and the origin of the Liouvillian skin effect for the general case through analytic methods.

\section{SII.~Analytic study of the Liouvillian}
In this section, we analytically solve the spectrum of the Liouvillian superoperator to elucidate the origin of the Liouvillian skin effect described in the previous section. 

First, we rearrange the Lindblad equation, in Eq.~(1) of the main text, into
\begin{equation}
\frac{\de\rho}{\de t}=-i[H_{\text{eff}},\rho]+\sum_{l,p}2L_{l,p}\rho L^{\dag}_{l,p},
\end{equation}
where the effective non-Hermitian Hamilton is
\begin{equation}
H_{\mathrm{eff}}=H-i \sum_{l,p}L^{\dag}_{l,p} L_{l,p}.
\end{equation}
We observe that the effective non-Hermitian Hamiltonian for the setup in Fig.~1 of the main text is block-diagonal.
This is because both the coherent Hamiltonian $H$ and the terms $-i \sum_{l,p}L^{\dag}_{l,p} L_{l,p}$ are block-diagonal with respect to the  subsystems shown in Fig.~1(a) of the main text.
We hence denote
\begin{equation}
H_{\text{eff}}=H_1\oplus H_2\oplus \cdots\oplus H_m,
\end{equation}
where $m$ represents the number of subsystems and each $H_i$ represents an individual subsystem with the Hilbert-space dimension $n_i$, with $\sum_{j=1}^m n_j = N$.

For our model in Fig.~1(b) of the main text, we find that the effective Hamilton is composed of two single-level systems $H_{1}=0$ and $H_{N/2+1}=\sum_{l=1}^{N/2-1}\omega_l-i\gamma_1$, and a series of $(N/2-1)$ two-level subsystems each described by the Hamiltonian
\begin{equation}
\begin{split}
H_{j}=&(\sum_{l=1}^{j}\omega_l)|g,j+1\rangle\langle g,j+1| +(\sum_{l=1}^{j-1}\omega_l-i\gamma_1) |e,j\rangle\langle e,j|+ \Omega_{j}(|e,j\rangle\langle g,j+1| + |g,j+1\rangle\langle e,j| ),
\end{split}\label{eq:subsytem}
\end{equation}
where $j=2,3,\cdots,N/2$.

Under the PBC, we observe that all $N/2$ subsystems in the effective Hamilton are two-level systems, given by the Hamiltonian in Eq.~(\ref{eq:subsytem}), but with
\begin{equation}
\begin{split}
H_{1}=&(\sum_{l=1}^{N/2-1}\omega_l-i\gamma_1) |e,N/2\rangle\langle e,N/2|+ \Omega_{N/2}(|e,N/2\rangle\langle g,1| + |g,1\rangle\langle e,N/2| ).
\end{split}
\end{equation}

\begin{figure*}[hbp!]
  \centering
  \includegraphics[width=13.5cm]{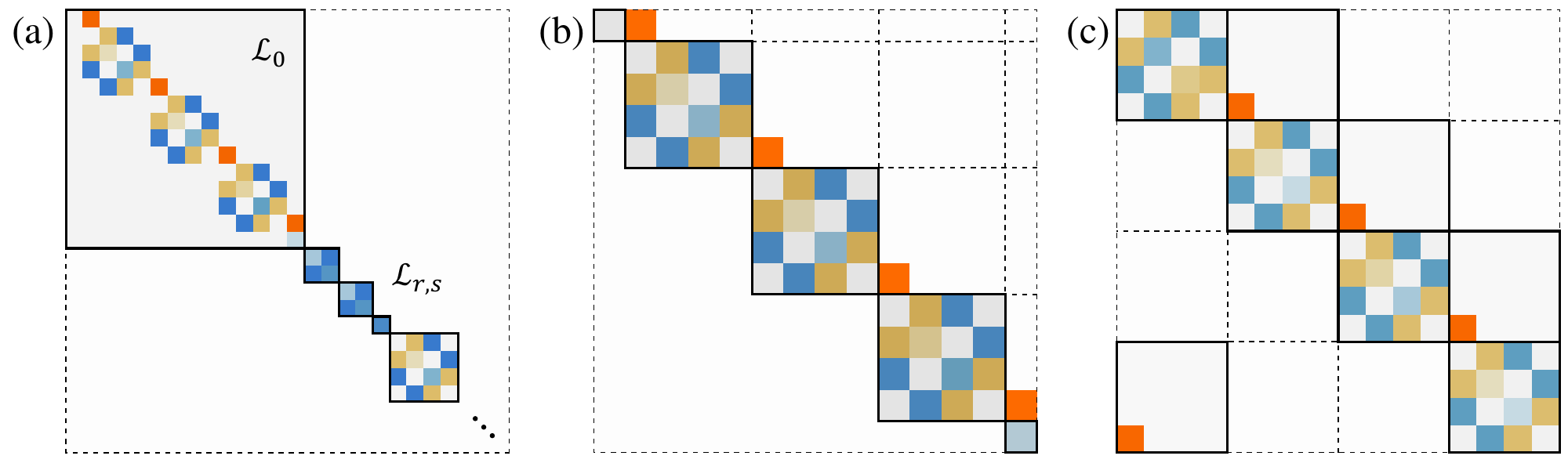}
  \caption{The matrix structure of the Liouvillian superoperator. (a) shows the block-diagonlized structure of the Liouvillian operator; (b) shows the block upper-triangular structure of $\mathcal{L}_0$ with OBC; (c) shows the block-circulant structure of $\mathcal{L}_0$ with PBC. Here blocks with different colors represent different matrix elements, and the orange blocks indicate the matrix elements from the recycling terms $\sum_{l,p}2L_{l,p}\rho L^{\dag}_{l,p}$. We fix the system size to  $N=8$. }
  \label{fig:figS3}
\end{figure*}

Additionally, we observe that the contribution from the recycling terms $\sum_{l,p}2L_{l,p}\rho L^{\dag}_{l,p}$ exists either between two adjacent subsystems, or within an individual subsystem (defined as $\mathcal{L}_0$ below). Hence the overall Liouvillian superoperator is also block-diagonal in its matrix form, as illustrated in Fig.~\ref{fig:figS3}(a). The large block with intra-block recycling-term contribution is
given by the Liouvillian
\begin{equation}
\mathcal{L}_0=-i\sum^m_{j=1}(H_j\otimes \mathcal{I}_{n_j}-\mathcal{I}_{n_j}\otimes H_{j})+\sum_{l,p}2L_{l,p}\rho L^{\dag}_{l,p},\label{L0}
\end{equation}
with the dimension $\sum^m_{j=1}n_j^2$, and $n_j$ represents the dimension of $H_j$.
Other blocks are given by
\begin{equation}
\mathcal{L}_{r,s}=-i(H_s\otimes \mathcal{I}_{n_r}-\mathcal{I}_{n_s}\otimes H_{r}),\label{Lrs}
\end{equation}
with dimensions $n_r n_s$, where $r,s=1,2,\cdots,m$ and $r\neq s$.  
Here $n_r$ and $n_s$ represent the dimensions of $H_r$ and $H_s$, respectively.
According to Eq.~(\ref{eq:subsytem}), $n_r$ and $n_s$ take the values of either 1 or 2, and $\mathcal{I}_{n}$ is the identity matrix with the dimension $n$.

Due to the block-diagonal structure of the Liouvillian, its eigenspectrum is analytically solvable by diagonalizing $\mathcal{L}_0$ and $\mathcal{L}_{r,s}$, respectively.
Specifically, in our model, since the dimensions of $H_r$ and $H_s$ are less than or equal to 2, the dimension of any given $\mathcal{L}_{r,s}$ is less than or equal to 4. And $\mathcal{L}_0$ is a special matrix that is easy to diagonalize.

We first study the case with OBC.
In this case, we observe that the dissipation between two adjacent subsystems is directional, which makes $\mathcal{L}_0$ a block upper-triangular matrix, as illustrated in Fig~\ref{fig:figS3}(b). The eigenspectrum of $\mathcal{L}_0$ is then the union of the spectra of the diagonal blocks. In the presence of translational symmetry with $\omega_l = 0$ and $\Omega_l = \Omega$, diagonal blocks of $\mathcal{L}_0$ are invariant with increasing system size.
Consequently, the Liouvillian gap remains constant as the system size changes, consistent with discussions in the previous section.
Furthermore, due to the block upper-triangular structure of $\mathcal{L}_0$, some eigenvectors from $\mathcal{L}_0$ are localized within the subsystems near the boundaries of the entire state space. As we detail in the next section, such a localization persists even as the translational symmetry is broken (for general values of $\omega_l$ and/or $\Omega_l$). Further, we provide the detailed calculation of the Liouvillian spectrum under OBC in Section SIV.

Under the PBC, when $\omega_l = 0$ and $\Omega_l = \Omega$, due to the translational symmetry, $\mathcal{L}_0$ forms a block-circulant matrix, illustrated in Fig~\ref{fig:figS3}(c).
We can thus visualize it as a four-band non-Hermitian one-dimensional lattice model with PBC.
The eigenspectrum is analytically solvable, and we find that the Liouvillian gap approaches zero when the system size tends to infinity. We provide more detail in Section SV.

Therefore, the Liouvillian skin effect observed in the maintext mathematically originates from the difference in $\mathcal{L}_0$ under different boundary conditions. Physically, the Liouvillian skin effect in our system arises from the divisibility of the effective Hamilton and the non-reciprocal recycling terms.
This phenomenon is analogous to the non-Hermitian skin effect observed in non-Hermitian lattice models.
Finally, we remark that our discussions here can be generalized to generic optical pumping setups illustrated in Fig.~1(a) of the main text.

\section{SIII.~The block upper-triangular matrix}
Here we discuss the eigen problem of a block upper-triangular matrix $M$ with
\begin{equation}
M=\begin{pmatrix}
A_{1,1}&A_{1,2}&A_{1,3}&\cdots&A_{1,m}\\
0 & A_{2,2}& A_{2,3}&\cdots&A_{2,m}\\
0 & 0&A_{3,3}& \cdots&\vdots\\
\vdots&\vdots&\vdots&\ddots&\vdots\\
0 & 0&0& \cdots&A_{m,m}\\
\end{pmatrix},
\end{equation}
where $A_{i,j}$ are matrices with dimensions $n_i\times n_j$, respectively.

We observe that the eigenvalues of the block upper-triangular matrix $M$ is the union of the spectra of the diagonal blocks $A_{i,i}$. In the following, we prove it by induction.

First, the conclusion obviously holds for $m=1$.
Then, assuming the statement is valid for $m=l$, we will show below that it also holds for $m=l+1$.
To simplify discussions, we set
\begin{align}
M_l&=\begin{pmatrix}
A_{1,1}&A_{1,2}&A_{1,3}&\cdots&A_{1,l}\\
0 & A_{2,2}& A_{2,3}&\cdots&A_{2,l}\\
0 & 0&A_{3,3}& \cdots&\vdots\\
\vdots&\vdots&\vdots&\ddots&\vdots\\
0 & 0&0& \cdots&A_{l,l}\\
\end{pmatrix},\\
C_l&=\begin{pmatrix}
A_{1,l+1}\\
A_{2,l+1}\\
\vdots\\
A_{l,l+1}\\
\end{pmatrix}.
\end{align}
For $m=l+1$, we have
\begin{align}
M_{l+1}=\begin{pmatrix}
M_l&C_l\\
0&A_{l+1,l+1}
\end{pmatrix}.
\end{align}

Any eigenvalue of $M_l$ is thus also an eigenvalue of $M_{l+1}$. Specifically, for any given eigenvalue $\alpha$ and the corresponding eigenvector $\ket{\psi}$ of $M_l$, we have
\begin{equation}
\begin{pmatrix}
M_l&C_l\\
0&A_{l+1,l+1}
\end{pmatrix}
\begin{pmatrix}
\ket{\psi}\\
0
\end{pmatrix}=\begin{pmatrix}
M_l\ket{\psi}\\
0
\end{pmatrix}=\alpha\begin{pmatrix}
\ket{\psi}\\
0
\end{pmatrix}.
\end{equation}

We then show that any eigenvalue of $A_{l+1,l+1}$ is also an eigenvalue of $M_{l+1}$. For that purpose, we
focus on a given eigenvalue $a^{l+1}$ and the corresponding eigenstate  $\ket{\psi^{l+1}}$ of $A_{l+1,l+1}$.
If $a^{l+1}$ is also an eigenvalue of $M_l$, with the corresponding eigenstate $\ket{\phi_0}$, we have
\begin{equation}
\begin{split}
\begin{pmatrix}
M_l&C_l\\
0&A_{l+1,l+1}
\end{pmatrix}
\begin{pmatrix}
\ket{\phi_0}\\0
\end{pmatrix}&=\begin{pmatrix}
M_l\ket{\phi_0}\\
0
\end{pmatrix}=a^{l+1}\begin{pmatrix}
\ket{\phi_0}\\
0
\end{pmatrix}
\end{split}.
\end{equation}

Otherwise, we have
\begin{equation}
\begin{split}
\begin{pmatrix}
M_l&C_l\\
0&A_{l+1,l+1}
\end{pmatrix}
\begin{pmatrix}
\ket{\phi}\\
\ket{\psi^{l+1}}
\end{pmatrix}&=\begin{pmatrix}
M_l\ket{\phi}+C_l\ket{\psi^{l+1}}\\
A_{l+1,l+1}\ket{\psi^{l+1}}
\end{pmatrix}=\begin{pmatrix}
M_l\ket{\phi}+C_l\ket{\psi^{l+1}}\\
a^{l+1}\ket{\psi^{l+1}}
\end{pmatrix}
\end{split},\label{key11}
\end{equation}
where $\ket{\phi}$ is an unknown state. We set
\begin{equation}
M_l\ket{\phi}-a^{l+1}\ket{\phi}=(M_l-a^{l+1}\mathcal{I})\ket{\phi}=C_l\ket{\psi^{l+1}},\label{key12}
\end{equation}
where $\mathcal{I}$ is an identity matrix.
Since $a^{l+1}$ is not an eigenvalue of $M_l$, $(M_l-a^{l+1}\mathcal{I})$ is reversible, and  Eq.~\ref{key12} must have nontrivial solutions. In other words, we can always find $\ket{\phi}$ such that Eq.~\ref{key11} is satisfied, yielding the right eigenstate of $M_{l+1}$.

In summary, the eigenvalue of $M_l$ and $A_{l+1,l+1}$ are the eigenvalues of $M_{l+1}$. In other words, our statement is also valid for $m=l+1$. We have therefore proved our statement by induction, that
the eigenvalues of the block upper-triangular matrix $M$ is the union of the spectra of the diagonal blocks $A_{i,i}$.

Furthermore, we notice that the right eigenstates of the block upper-triangular matrix are usually localized in the Hilbert space. Here we provide a simple explanation. We set
\begin{equation}
\begin{split}
&M_k=\begin{pmatrix}
A_{1,1}&A_{1,2}&A_{1,3}&\cdots&A_{1,k}\\
0 & A_{2,2}& A_{2,3}&\cdots&A_{2,k}\\
0 & 0&A_{3,3}& \cdots&\vdots\\
\vdots&\vdots&\vdots&\ddots&\vdots\\
0 & 0&0& \cdots&A_{k,k}\\
\end{pmatrix},\\&D_k=\begin{pmatrix}
A_{1,l+1}&\cdots&A_{1,l+1}\\
A_{2,l+1}&\cdots&A_{1,l+1}\\
\vdots\\
A_{l,l+1}&\cdots&A_{1,l+1}\\
\end{pmatrix},\\
&M'_k=\begin{pmatrix}
A_{k+1,k+1}&A_{k+1,k+2}&A_{k+1,k+3}&\cdots&A_{k+1,m}\\
0 & A_{k+2,k+2}& A_{k+2,k+3}&\cdots&A_{k+2,m}\\
0 & 0&A_{k+3,k+3}& \cdots&\vdots\\
\vdots&\vdots&\vdots&\ddots&\vdots\\
0 & 0&0& \cdots&A_{m,m}\\
\end{pmatrix}
\end{split}.
\end{equation}
the entire matrix follows
\begin{equation}
M_{l}=\begin{pmatrix}
M_k&D_k\\
0&M'_k
\end{pmatrix}.
\end{equation}

Following the previous discussion, for any $k$, the eigenvalues of $M_k$ are also the eigenvalues of $M_l$ and the corresponding right eigenstate is equal to 0 in the space of $M'_k$. Therefore, these right eigenstates are localized within the space of $M_k$.

\section{SIV.~OBC Liouvillian spectrum}
Here we analyze the Liouvillian spectrum of our model for any $\Omega_l$ and $\omega_l$ under the OBC.
As discussed in Sec.~III, the Liouvillian operator $\mathcal{L}$ is composed of $\mathcal{L}_0$ [defined in Eq.~(\ref{L0})], and $\mathcal{L}_{r,s}$ [defined as Eq.~(\ref{Lrs})].
We denote the eigenvalues of $\mathcal{L}_{r,s}$ as $-i(\epsilon_{l,r}+\epsilon_{m,s}^*)$,
where $\epsilon_{l,r}$ is the $l$-th eigenvalues of $H_r$, $\epsilon_{m,s}^*$ is the complex conjugate of the $m$-th eigenvalue of $H_s$, and $1\le i \le n_r,1\le j \le n_s$. The number of eigenvalues of $\mathcal{L}_{r,s}$ is then $N^2-2N+2$ when the system size is N.

Due to the block upper-triangular structure of $\mathcal{L}_0$, the spectrum of $\mathcal{L}_0$ is the union of the spectra of the diagonal blocks. The diagonal blocks of $\mathcal{L}_0$ include two one-dimensional blocks with diagonals of $0$ and $-\gamma_1$, as well as $N/2-1$ four-dimensional blocks
in the following form
\begin{equation}B_l=
\begin{pmatrix}
-\gamma&i\Omega_l&-i\Omega_l&\gamma_0\\
i\Omega_l&-\frac{1}{2}\gamma+i\omega_l&0&-i\Omega_l\\
-i\Omega_l&0&-\frac{1}{2}\gamma+i\omega_l&i\Omega_l\\
0&-i\Omega_l&i\Omega_l&0
\end{pmatrix},
\end{equation}
where $\gamma=\gamma_0 + \gamma_1$. The number of eigenvalues of $\mathcal{L}_{0}$ is $2N-2$ when the system size is $N$.

When $\Omega_l=\Omega$ and $\omega_l=\omega$, due to the translational symmetry, the whole Liouvillian spectrum is invariant (as we discussed in the main text) but the degeneracy of the eigenmodes increases with increasing system size.

When $\Omega_l=\sqrt{l}\Omega$ and $\omega_l=\omega$, we find numerically that the eigenvalue whose real component is closest to 0 always comes from the first four-dimensional diagonal block of $\mathcal{L}_0$, which comes from the $l=1$ sector. Therefore, in this case, the Liouvillian gap is invariant with increasing system size.

\section{SV.~Liouvillian gap}
Here we provide analytic expressions for the Liouvillian gap in the main text.

Under the OBC, when $\gamma_0=0$, $\omega_l = 0$ and $\Omega_l=\Omega$, the Liouvillian gap follows
\begin{equation}
\Delta_{OBC}=\begin{cases}
\frac{1}{4}(\gamma_1-\sqrt{\gamma_1^2-16\Omega^2}), & \text{for}\,\,\frac{\Omega}{\gamma_1}<\frac{1}{4},\\
\frac{1}{4}\gamma_1,&\text{for}\,\,\frac{\Omega}{\gamma_1}\ge\frac{1}{4}.
\end{cases}
\end{equation}
If we consider $\omega_l=\omega$, the Liouvillian gap becomes
\begin{equation}
\Delta=\frac{1}{4}(\gamma_1-\text{Im}\left[\sqrt{\gamma_1^2-4\omega^2-4i\omega\gamma_1-16\Omega^2}\right ]).
\end{equation}

Under the PBC, when $\omega_l = 0$, $\Omega_l = \Omega$ and $\gamma_0=0$, we regard $\mathcal{L}_0$ as a four-band, one-dimensional lattice. Due to the lattice translational symmetry, its Hamilton can be written in the $k$ space as
\begin{equation}
H_4=\begin{pmatrix}
-\gamma_1&i\Omega&-i\Omega&\gamma_1 e^{ik}\\
i\Omega&-\frac{1}{2}\gamma_1&0&-i\Omega\\
-i\Omega&0&-\frac{1}{2}\gamma_1&i\Omega\\
0&-i\Omega&i\Omega&0
\end{pmatrix}.
\label{pbc4band}
\end{equation}
Here $k=m\pi/N(m=1,2,\cdots,N/2)$ is the lattice momentum. The Liouvillian gap can be calculated from  the spectrum of
Eq.~\ref{pbc4band}. 

We then calculate the Liouvillian gap after introducing the decay term $L_{l,1}$, under the OBC and with $\omega_l = \omega$ and $\Omega_l = \Omega$.

In order to derive the Liouvillian gap, we need to calculate the spectrum of each diagonal block in $\mathcal{L}_0$, as well as the blocks $\mathcal{L}_{r,s}$. When $\omega\neq0$, the expression of the Liouvillian gap is extremely complicated. However, we notice that the Liouvillian gap consistently decreases with increasing energy interval $\omega$. Thus, we calculate the Liouvillian gap for $\omega=0$ for an upper bound, which is given by
\begin{equation}
\Delta=\begin{cases}
\frac{1}{4}(\gamma_0+\gamma_1-\sqrt{(\gamma_0+\gamma_1)^2-16\Omega^2}), &\text{for}\,\,4\Omega<\gamma_0+\gamma_1,527 \gamma _0^2+575 \gamma _1^2+1166 \gamma _0 \gamma _1>9216 \Omega ^2,\\
\frac{1}{4}(\gamma_0+\gamma_1),&\text{for}\,\,4\Omega\ge\gamma_0+\gamma_1,64\Omega^2(\gamma_1-\gamma_0)>3(\gamma_1+\gamma_0)^3,\\
\frac{\gamma _0+\gamma _1}{2} -\frac{\sqrt[3]{72 \gamma _0 \Omega ^2+\frac{1}{3} \sqrt{46656 \gamma _0^2 \Omega ^4+\left(-3 \gamma _0^2-3 \gamma _1^2-6 \gamma _0 \gamma _2+48 \Omega ^2\right){}^3}}}{2\ 3^{2/3}}\\
-\frac{\gamma _0^2+\gamma _1^2+2 \gamma _0 \gamma _1-16 \Omega ^2}{2 \sqrt[3]{3} \sqrt[3]{72 \gamma _0 \Omega ^2+\frac{1}{3} \sqrt{46656 \gamma _0^2 \Omega ^4+\left(-3 \gamma _0^2-3 \gamma _1^2-6 \gamma _0 \gamma _1+48 \Omega ^2\right){}^3}}},&\text{otherwise}.
\end{cases}
\label{maxdelta}
\end{equation}

According to Eq.~(\ref{maxdelta}), for certain $\Omega$ and $\gamma_1$ , we have
\begin{equation}
\gamma_{1,max}=\begin{cases}
0,&\text{for}\,\,4\Omega<\gamma_1,\\
4\Omega-\gamma_1,&\text{for}\,\,4\Omega\ge\gamma_1\ge\frac{7}{2}\Omega,\\
-\frac{4}{3} \sqrt[3]{\sqrt{81 \gamma _1^2 \Omega ^4+64 \Omega ^6}-9 \gamma _1 \Omega ^2}+\frac{16 \Omega ^2}{3 \sqrt[3]{\sqrt{81 \gamma _1^2 \Omega ^4+64 \Omega ^6}-9 \gamma _1 \Omega ^2}}-\gamma _1,&\text{for}\,\,\gamma_1<\frac{7}{2}\Omega.
\end{cases}
\end{equation}

The maximum Liouvillian gap is therefore
\begin{equation}
\Delta_{max}(\gamma_1,\Omega)=\begin{cases}
\frac{1}{4}(\gamma_1-\sqrt{\gamma_1^2-16\Omega^2}), & \text{for}\,\,4\Omega<\gamma_1,\\
\Omega,&\text{for}\,\,4\Omega\ge\gamma_1\ge\frac{7}{2}\Omega,\\
-\frac{1}{3}  \sqrt[3]{\sqrt{81 \gamma _1^2 \Omega ^4+64 \Omega ^6}-9 \gamma _1 \Omega ^2}+\frac{4 \Omega ^2}{3 \sqrt[3]{\sqrt{81 \gamma _1^2 \Omega ^4+64 \Omega ^6}-9 \gamma _1 \Omega ^2}},&\text{for}\,\,\gamma_1<\frac{7}{2}\Omega.
\end{cases}
\end{equation}

When $\Omega\to\infty$, the maximum possible Liouvillian gap is $\gamma_1/2$.

